\documentclass[letterpaper,12pt]{article}
\usepackage{cite}
\usepackage[all]{xy}
\usepackage[dvips]{graphicx}
\usepackage{graphpap}
\usepackage{ifthen}
\usepackage{enumerate}
\usepackage{amssymb}
\usepackage{mathrsfs}
\usepackage[errorshow]{tracefnt}
\usepackage{fancybox}
\usepackage{amsfonts}
\usepackage{amsmath}
\usepackage{amssymb}
\usepackage{multirow}
\usepackage{array}
\usepackage{mathtools}
\usepackage{soul}
\usepackage{pict2e}

\makeatletter
\newcommand{\thickhline}{%
    \noalign {\ifnum 0=`}\fi \hrule height 1pt
    \futurelet \reserved@a \@xhline
}
\newcolumntype{'}{@{\hskip\tabcolsep\vrule width 1pt\hskip\tabcolsep}}
\makeatother
\newcolumntype{"}{@{\hskip\tabcolsep\vrule width 1.5pt\hskip\tabcolsep}}
\makeatother

\renewcommand{\cal}{\mathscr}
\renewcommand{\mathcal}{\mathscr}

\newcommand{\ket}[1]{| #1 \rangle}

\newcommand{\cA}{\mathcal{A}}
\newcommand{\cB}{\mathcal{B}}

\newcommand{\cD}{\mathcal{D}}

\newcommand{\bvee}{\text{\large{$\vee$}}}
\newcommand{\bwedge}{\text{\large{$\wedge$}}}

\def\lb{{\lambda}}
\def\gf{{\mathfrak g}}
\def\af{{\mathfrak a}}
\def\ie{{\it i.e.}}
\def\eg{{\it e.g.}}
\newcommand\BB{{\cal B}}
\newcommand\ZZ{{\cal Z}}

\def\boxit#1{\vbox{\hrule\hbox{\vrule\kern3pt
             \vbox{\kern3pt#1\kern3pt}\kern3pt\vrule}\hrule}}

\newcommand{\beq}{\begin{equation}}
\newcommand{\beqn}{\begin{equation*}}
\newcommand{\eeq}{\end{equation}}
\newcommand{\eeqn}{\end{equation*}}
\newcommand{\beqa}{\begin{eqnarray}}
\newcommand{\beqan}{\begin{eqnarray*}}
\newcommand{\eeqa}{\end{eqnarray}}
\newcommand{\eeqan}{\end{eqnarray*}}
\newcommand{\bdm}{\begin{displaymath}}
\newcommand{\edm}{\end{displaymath}}

\newcommand{\la}{\langle}
\newcommand{\ra}{\rangle}

\newcommand{\ba}{\begin{array}}
\newcommand{\ea}{\end{array}}

\newcommand\nn{\nonumber}

\newcommand\benu{\begin{enumerate}}
\newcommand\eenu{\end{enumerate}}
\newcommand\bit{\begin{itemize}}
\newcommand\eit{\end{itemize}}

\def\der'{\mathfrak{der}'\,}
\def\der{\mathfrak{der}\,}
\def\str'{\mathfrak{str}'\,}
\def\str{\mathfrak{str}\,}

\def\g{\mathfrak{g}}

\newcommand{\ad}{\text{ad}\,}

\RequirePackage{hyperref}

\numberwithin{equation}{section}

\begin{document}

\frenchspacing

%\vskip-10pt
%\hfill {\tt brst8.tex} \\
%\vskip-10pt
%\hfill {\tt \today} \\
\vskip-10pt
\hfill {\tt MI-TH-1508} \\
\vskip-10pt
\hfill {\tt Gothenburg preprint}\\
\vskip-10pt
\hfill {\tt arXiv:1503.06215}

\pagestyle{empty}

\vspace*{1.5cm}

\begin{center}
\noindent
{\LARGE {\sf \textbf{Superalgebras, constraints}}}\\
%\noindent
\vspace{.3cm}
\noindent
{\LARGE {\sf \textbf{and partition functions}}}\\
\vspace{.3cm}

\renewcommand{\thefootnote}{\fnsymbol{footnote}}

\vskip 1truecm
\noindent
{\large {\sf \textbf{Martin Cederwall}
}}\\
\vskip .5truecm
        {\it Department of Fundamental Physics\\
        {Chalmers University of Technology\\ SE-412 96 G\"oteborg, Sweden}\\[3mm]}
        {\tt martin.cederwall@chalmers.se} \\
\vskip 1truecm
\noindent
{\large {\sf \textbf{Jakob Palmkvist}
}}\\
\vskip .5truecm
        {\it 
        {
        %George and Cynthia Woods 
        Mitchell Institute for Fundamental Physics and Astronomy\\
        Texas A\&M University\\ College Station, TX 77843, USA}\\[3mm]}
        {\tt jakobpalmkvist@tamu.edu} \\
\end{center}

\vskip 1cm

\centerline{\sf \textbf{
Abstract}}
\vskip .2cm

\noindent
We consider Borcherds
superalgebras obtained from semisimple finite-dimen\-sional Lie
algebras by adding 
an odd null root to the simple roots. The additional Serre relations can be
expressed in a covariant way. The spectrum of generators at positive
levels are associated to partition functions for a certain set of
constrained bosonic variables, the constraints on which are
complementary to the Serre relations in the symmetric product. 
We give some examples, focusing on superalgebras related to pure spinors,
exceptional geometry and tensor hierarchies, of how construction of the 
content of the algebra at arbitrary
levels is simplified.

\newpage

\pagestyle{plain}

\tableofcontents

\section{Introduction}\label{Introduction}

It is often useful in physics to describe a spectrum of states that
appear at various integer levels by means of an associated partition
function, especially if the spectrum is infinite. If the states at
each level transform in a representation of a Lie algebra,
the spectrum of representations may also be obtained from an
extended (possibly infinite-dimensional) algebra by a level
decomposition. In the present paper we will relate these two
approaches to each other, and also to a third important tool in
physics: the BRST treatment of reducible constraints.

Our main example is the spectrum of dynamical forms in $D$-dimensional
maximal supergravity, which transform in representations of the
U-duality group\footnote{Throughout the paper, we use the
notation $E_n$ for the split real form,
and also for the corresponding Lie algebra. All arguments are however
equally valid for the complex Lie algebras.} $E_{n}$, where $n=11-D$. Remarkably, these
representations form a Lie superalgebra, which can be extended to an
infinite-dimensional Borcherds superalgebra
\cite{Cremmer:1998px,HenryLabordere:2002dk,Henneaux:2010ys}. Decomposing
it with respect to the $E_{n}$ subalgebra gives back the spectrum
of dynamical forms at the positive levels, and also precisely the
additional non-dynamical forms allowed by supersymmetry, first determined for $D=10$ in
refs.~\cite{Bergshoeff:2005ac,Bergshoeff:2006qw,Bergshoeff:2010mv}.  The
consistency with supersymmetry was shown in
refs.~\cite{Greitz:2011vh,Greitz:2011da,Greitz:2012vp,Howe:2015hpa} using
a superspace formulation, generalising bosonic forms to superforms
with arbitrary high degrees.  However, already in the restriction to
the bosonic sector, the wedge product in the differential algebra,
with even and odd forms, naturally gives rise to a superalgebra
structure.  Up to an arbitrary level $p$ the representations can be
also obtained from a level decomposition of the Kac--Moody algebra
$E_{n+p}$
\cite{Henneaux:2010ys,Palmkvist:2011vz,Palmkvist:2012nc,Howe:2015hpa}.
This generalises results for $E_{11}$, which as a special case
contains the form spectrum up to $p=D$
\cite{Riccioni:2007au,Bergshoeff:2007qi,Bergshoeff:2007vb}. However, $E_{11}$ is not
enough to accommodate forms with higher degrees, and a rendition of
{\it all} the representations coming from the Borcherds superalgebra
would require a consideration of the infinite-rank algebra $E_\infty$.

With a few exceptions, the level decomposition of the Borcherds
superalgebra furthermore agrees with the tensor hierarchy of form potentials,
field strengths and gauge parameters that arises in the
embedding tensor approach to gauged supergravity
\cite{deWit:2003hr,deWit:2005hv,deWit:2008ta,deWit:2008gc}. The tensor hierarchy
can be continued to infinity, but misses some of the representations
coming from the Borcherds superalgebra. Perfect agreement is instead
given by a tensor hierarchy algebra, where the embedding tensor is
interpreted as an element at level minus one \cite{Palmkvist:2013vya}.
Using this algebra all the Bianchi identities and gauge
transformations for the gauged theory can be derived in a simple way
\cite{Greitz:2013pua,Howe:2015hpa}. This demonstrates the efficiency of organising
representations into a level decomposition of a Lie (super)algebra.

Yet another context where the same infinite sequence of
representations appears, and where it cannot be truncated, is 
exceptional geometry. 
The exceptional (generalised) diffeomorphisms have infinite
reducibility, and
the sequence arises as the tower of ghosts for ghosts, describing this
reducibility \cite{Berman:2012vc}. The connection to partition
functions of constrained objects, of which pure spinors
\cite{Berkovits:2005hy} is one example, was conjectured already in
ref.~\cite{Berman:2012vc}, and used there to correctly regularise the
infinite sums arising when counting the degrees of freedom. 
The same representations occur for tensor fields in exceptional
geometry \cite{Cederwall:2013naa}, and in the tensor hierarchies
considered in \cite{Hohm:2013jma,Hohm:2013vpa,Hohm:2013uia,Hohm:2014fxa}.
The somewhat heuristic approach of ref.~\cite{Berman:2012vc}
provided one of the motivations for the present
investigation, which puts the correspondence between the algebra and
the constrained objects on a firmer footing.

For $3 \leq D \leq 8$ the U-duality algebra $E_n$ is extended to the
infinite-dimensional Borcherds superalgebra $\BB$ by adding an odd
null root $\beta_0$ to the simple roots of $E_n$. This is the special
case that we focus on in this paper, with $E_n$ generalised to any
semisimple finite-dimensional Lie algebra $\mathfrak g$.
The inner products of $\beta_0$ with the simple roots of $\mathfrak g$
are assumed to be such that the Serre relations of the Borcherds
superalgebra are at most quadratic in the odd Chevalley generators $e_0,f_0$
corresponding to $\beta_0$.  Denoting the representation at level $p$
by $R_p$, the Serre relations quadratic in $e_0$ (say) thus belong to
a representation of $\mathfrak g$ contained in the symmetric tensor
product of $R_1$ with itself, with $R_2$
as its complement\footnote{With the complement of a representation
$R$ in another $R'$ we mean the quotient $R'-R$.}. It
generates an ideal of the free Lie superalgebra generated by $R_1$,
and at each level $p\geq2$, the representation $R_p$ is the complement
that is left when this ideal is factored out.  A recursive study of
the ideal thus gives all information about the representations $R_p$
at any level $p$.  In this paper, we will show that the
representations $R_p$ alternatively, and often more directly, can be
determined from the partition function for a bosonic object $\lb$ in
$R_1$, subject to the constraint $\lb^2|_{R_2}=0$. As our main result,
we will show that {\it this partition function is the inverse of the
partition function for the universal enveloping algebra of $\BB_+$,}
the subalgebra of $\BB$ at positive levels.

The paper is organised as follows. In Section 2 we describe in more
detail the Lie superalgebras that we consider, and how they are
constructed from the Chevalley generators and the Serre relations. In
Section 3 we introduce the partition functions that we use in Section
4 to state our results and give them an interpretation in terms of a
BRST operator. The argument of Section 4 
corresponds roughly to the heuristic argument of
ref. \cite{Berman:2012vc}.
We then prove the result in Section 5 using the denominator
formula for Borcherds superalgebras. Section 6 addresses the question
why the method is not 
applicable to Lie algebras (extensions by an additional even real
root). In Section 7 we present and discuss some examples.

%%%%%%%%%%%%%%%%%%%%%%%%%%%%%%%%%%%%%%%%%%%%%%%%%%%%%%%%%%%%%%%%%%%%%%%%%%%%

\section{The superalgebras}\label{SuperAlgebraSub}

%%%%%%%%%%%%%%%%%%%%%%%%%%%%%%%%%%%%%%%%%%%%%%%%%%%%%%%%%%%%%%%%%%%%%%%%%%%%

\renewcommand{\cal}{\mathscr}

Let $\g$ be a semisimple finite-dimensional Lie algebra of rank $r$
with simple roots $\alpha_i$ ($i=1,\ldots,r$). We recall that they
form a basis of a euclidean space, and from their mutual inner
products we get the Cartan matrix $a_{ij}$ of $\g$ by
\begin{align}
a_{ij}=(\alpha_j,\alpha_i{}^\vee)
=2\frac{(\alpha_j,\alpha_i)}{(\alpha_i,\alpha_i)}\ ,
\end{align}
where $\alpha_i{}^\vee=2\alpha_i/(\alpha_i,\alpha_i)$ is the coroot of
$\alpha_i$.

The construction of a Lie algebra from a basis of simple roots can be generalised to
inner product spaces which are not necessarily
euclidean, and even from Lie algebras to Lie superalgebras.
Semisimple finite-dimensional Lie algebras are then generalised
to Kac-Moody (super)algebras, which in turn are generalised further to Borcherds (super)algebras.
Thus Borcherds superalgebras
is a very general concept, but in this paper we only consider the special cases described below,
motivated by their simplicity and by their appearance in the
examples that we will study in Section 7. 
We refer the reader to refs.~\cite{Kac77B,Kac,Wakimoto,Ray} for more general definitions
and other details about Borcherds and Kac-Moody (super)algebras.

The Borcherds superalgebras that we consider 
are infinite-dimensional superextensions of semisimple
finite-dimensional Lie algebras, 
obtained by adding an odd null root to the simple roots. Let $\BB$ be
such an extension of $\g$, with simple roots 
$\beta_I$ ($I=0,1,\ldots,r$). Thus $\beta_0$ is odd and null,
$(\beta_0,\beta_0)=0$, whereas 
$\beta_i=\alpha_i$ are even and real, $(\beta_i,\beta_i)>0$.
The Cartan matrix $B_{IJ}$ of $\BB$ is obtained from $a_{ij}$ by
adding an extra column 
\begin{align}
B_{i0}=(\beta_0,\beta_i{}^\vee)=2\frac{(\beta_0,\beta_i)}{(\beta_i,\beta_i)}
\end{align}
and an extra row $B_{0I}=(\beta_I,\beta_0)$, including the diagonal entry
$B_{00}=(\beta_0,\beta_0)=0$. The additional off-diagonal entries 
$B_{i0}$ are required to be non-positive integers, like  
$B_{ij}=a_{ij}$ for $i\neq j$.
We assume furthermore that $B_{IJ}$ is non-degenerate, and for each
$i=1,\ldots,r$, either $B_{0i}=0$ or $B_{0i}=-1$. 

Adding an extra column and row to the Cartan matrix of $\g$
corresponds to adding an extra node to the Dynkin diagram of $\g$,
connected with 
$|B_{i0}|$ lines to 
node $i$.
Following ref. \cite{Kac77B} we indicate that $\beta_0$ is both null and odd by
painting the corresponding node ``grey'' (which means that it looks
like $\otimes$), 
and let the other nodes, representing real
even roots, be white.
For example, the Dynkin diagram\\
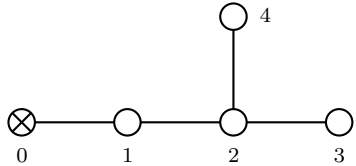
\begin{center}
\begin{picture} (130,60)(55,-15)      
\thicklines
\put(60,0){\circle{10}}
\put(56.5,-3.5){\line(1,1){7}}
\put(56.5,3.5){\line(1,-1){7}}
\put(100,0){\circle{10}}
\put(140,0){\circle{10}}
\put(180,0){\circle{10}}
\put(140,40){\circle{10}}
\put(65,0){\line(1,0){30}}
\put(105,0){\line(1,0){30}}
\put(145,0){\line(1,0){30}}
\put(140,5){\line(0,1){30}}
\put(58,-15){\scriptsize{$0$}}
\put(98,-15){\scriptsize{$1$}}
\put(138,-15){\scriptsize{$2$}}
\put(178,-15){\scriptsize{$3$}}
\put(150,38){\scriptsize{$4$}}
\end{picture}
\end{center} 
corresponds to the Cartan matrix
\begin{align}B_{IJ}=
\begin{pmatrix}
0 & -1 & 0 & 0&0\\
-1&2&-1&0&0\\
0&-1&2&-1&-1\\
0&0&-1&2&0\\
0&0&-1&0&2
\end{pmatrix}\ .
\end{align}
We will come back to this algebra, among other examples, in Section
\ref{ExamplesSect}. 

To each simple root $\beta_I$ of $\BB$ we associate Chevalley
generators $e_I$, $f_I$ and $h_I$, 
and $\BB$ is then defined as the Lie superalgebra generated by these
elements (of which $e_0$ and $f_0$ are odd and the others even) 
modulo the Chevalley relations
\begin{align} \label{chev-rel-1}
[h_I,e_J]&=B_{IJ}e_J\ , & 
[h_I,f_J]&=-B_{IJ}f_J\ , & [e_I,f_J\}&=\delta_{IJ}h_J\ ,
\end{align}
and the Serre relations 
\begin{align} \label{serre-rel1}
(\ad e_I)^{1-B_{IJ}} (e_J) = (\ad f_I)^{1-B_{IJ}} (f_J) &=0
\end{align}
for $I \neq J$. For $I=0$ the Serre relations (\ref{serre-rel1}) can
equivalently be replaced by 
\begin{align} \label{serre-rel2}
\{e_0,e_0\}=\{f_0,f_0\}=0\ ,
\end{align}
since, by the Jacobi identity,
\begin{align}
\tfrac12[\{e_0,e_0\},e_J]&=\{e_0,[e_0,e_J]\},&\tfrac12[\{f_0,f_0\},f_J]&=\{f_0,[f_0,f_J]\}\ ,
\end{align}
which gives
\begin{align}
[\{e_0,[e_0,e_J]\},f_J]&=
\{e_0,e_0\},&
[\{f_0,[f_0,f_J]\},e_J]&=
\{f_0,f_0\}
\end{align}
if $B_{J0}=-1$. Thus in this case the ideal generated by (\ref{serre-rel1})
is contained in the ideal generated by (\ref{serre-rel2}), and conversely.
If $B_{J0}=0$, there is already a redundance in (\ref{serre-rel1}) because of the antisymmetry
of the bracket, so replacing (\ref{serre-rel1}) by (\ref{serre-rel2}) in this case simply amounts to removing one of two equivalent
relations in (\ref{serre-rel1}).

For any integer $p$, let $\BB_p$ be the subspace of $\BB$ spanned by
all root vectors corresponding to roots $\beta=p\beta_0+\alpha$,
where $\alpha$ is a linear combination of the real simple roots
$\beta_i=\alpha_i$, and, if $p=0$, by the Cartan elements $h_I$. Since $\BB$ is
the direct sum of all these subspaces, and
$[\BB_p,\BB_q\}\subseteq\BB_{p+q}$, this decomposition is a
$\mathbb{Z}$-grading of $\BB$, leading to a level decomposition of its
adjoint representation under the subalgebra $\g \subset \BB_0$, with
$\BB_p$ consisting of a (maybe reducible)
representation\footnote{Following the physics terminology, we use the
  term ``representation'' also for the module of the representation,
  \ie, the vector space it acts on.} of $\g$ at level $p$.
We will throughout the paper denote this representation $R_p$.

Let $\BB_+$ and $\BB_-$ be the subalgebras of $\BB$ spanned by
elements at positive and negative levels, respectively, and let
$\tilde{\BB}_\pm$ be the free Lie superalgebra generated by
$\BB_{\pm1}$.  The Serre relations (\ref{serre-rel2}) generate an
ideal of $\tilde{\BB}$ which is the direct sum of two subalgebras
$\cD_\pm$, where $\cD_\pm\subset \tilde{\BB}_\pm$ (and is the
maximal such ideal). The Borcherds superalgebra $\BB$ is then obtained
by factoring out this ideal from $\tilde{\BB}$, and in particular
$\BB_+$ is obtained by factoring out $\cD_+$ from $\tilde{\BB}_+$, the
free Lie superalgebra generated by $\BB_1$.  The ideal $\cD_+$ of
$\tilde{\BB}$ is generated by the element $\{e_0,e_0\}$ at level two,
which is set to zero in one of the Serre relations (\ref{serre-rel2}).
However, considered as an ideal of $\tilde{\BB}_+$ only, it is
generated by {\it all} elements at level two in $\cD_+$, which are not
only $\{e_0,e_0\}$ but also those obtained from $\{e_0,e_0\}$ by
successively acting with $\g$. These elements form a representation 
$R_2{}^\perp$, which is the complement of $R_2$ in
$\vee^2 R_1$ (the symmetrised tensor product of $R_1$ with itself)
with a lowest weight vector $\{e_0,e_0\}$. It then follows from the
Chevalley relations that the Dynkin labels
of the lowest weight
of $R_2{}^\perp$ 
are given by $\lambda_i=2B_{i0}$.

Using a basis $E_M$ of $\BB_1$ (so that the index $M$ corresponds to
the representation $R_1$), 
we can summarize the above construction of $\BB_+$ by saying that it
is the Lie superalgebra generated by 
the odd elements $E_M$ modulo the 
``covariant Serre relations''
\begin{align}
\{E_M,E_N\}|_{R_2{}^\perp}=0\,.
\end{align}
Recursive use of these relations (and of course of the Jacobi
identity)  
gives complete information about
the representation $R_p$ at arbitrarily high levels $p$.
In Section \ref{brst-section} we will describe how 
this information can be efficiently encoded into partition functions,
which will be discussed next. 

%%%%%%%%%%%%%%%%%%%%%%%%%%%%%%%%%%%%%%%%%%%%%%%%%%%%%%%%%%%%%%%%%%%%%%%%%%%%

\section{Partition functions}\label{PartFuncSub}

%%%%%%%%%%%%%%%%%%%%%%%%%%%%%%%%%%%%%%%%%%%%%%%%%%%%%%%%%%%%%%%%%%%%%%%%%%%%

\renewcommand{\cal}{\mathscr}

The purpose of this section is to introduce and define notation
for the partition functions we use to state our results.

The partition functions we will consider count the number of bosonic
and fermi\-onic objects occurring with some $\mathbb Z$-weight, or
level, and some
additional quantum numbers. In an ``unrefined'' partition function,
only counting the number of states per level, the
presence of some set of $N$ linearly independent objects with weight
$p$ corresponds to a term
$\sigma^p N t^p$, where
$\sigma=1$ for bosons and $\sigma=-1$ for fermions.
This means we are really considering
partition functions twisted by fermion number, which of course has the
advantage that the partition function for a fermionic variable, or 
``creation operator'',
\begin{align}
Z^F(t)=1-t\,,
\end{align}
and that of a bosonic one,
\begin{align}
Z^B(t)=(1-t)^{-1}\,,
\end{align}
are each other's inverses.

More refined partition functions may be defined if
additional quantum numbers are available. In a typical case, a
variable or operator will transform in some representation $R$ of a Lie
algebra $\gf$. A refined partition function encodes
completely the representations of all states, and is a formal power series in
a variable $t$ (corresponding to the grading) with coefficients in
the unit ring of $\gf$-representations (under tensor product).

The basic examples are the refined partition functions for fermionic
and bosonic
creation operators in $R$:
\begin{align}
\ZZ^F_R(t)&=\bigoplus\limits_{p=0}^{|R|}(-t)^p\text{\large{$\wedge$}}^p R\,,
\nn\\
\ZZ^B_R(t)&=\bigoplus\limits_{p=0}^\infty t^p\,\text{\large{$\vee$}}^p R\,.
\label{ZZRFExpl}
\end{align}
Here, we use $\wedge$ and $\vee$ for antisymmetric and and symmetric
products, respectively, and $|R|$ denotes the dimension of a representation $R$.
These two functions are also the inverses of each other, when
multiplication is taken as the tensor product with the trivial representation
as the identity. This can be seen explicitly at any order in $t$ by
observing that the tensor product 
$(\wedge^p R)\otimes(\vee^q R)$ generically contains exactly the
plethysms described by the two different ``hook'' Young tableaux of 
$\mathfrak{sl}(|R|)$ obtained by gluing together the column and the row
describing the two factors.
One thus has 
\begin{align}
\ZZ^F_R(t)\otimes\ZZ^B_R(t)=1\,,
\end{align}
It is then reasonable to use the formal notation 
\begin{align}
\ZZ^F_R(t)&=(1-t)^R\,,\nn\\
\ZZ^B_R(t)&=(1-t)^{-R}\,.\label{ZZRFDef}
\end{align}
A fermion in $R$ can be seen as a boson in $-R$ and vice versa.
It is important to understand the notation of eq. (\ref{ZZRFDef}) as
the shorthand it is, with eq. (\ref{ZZRFExpl}) being its defining expression.

All considerations of the refined partition functions may also
be performed using characters, since they provide a ring homomorphism.
Writing the character of the representation $R$ as
$\chi(R)=\sum_{\mu\in\Lambda_R}e^\mu$, where $\Lambda_R$ is
the set of weights for $R$, counting weights with multiplicities
$m>1$ as $m$ ``distinct'' weights, we have
\begin{align}
\chi(\text{\large{$\wedge$}}^kR)=\sum_{\{\mu_1,\ldots,\mu_k\}}
e^{\mu_1+\cdots+\mu_k}\,,
\end{align}
where the sum is over sets of distinct (in the sense above)
weights in $\Lambda_R$. Thus,
\begin{align}
\chi^F_R(t)\equiv\chi(\ZZ^F_R(t))=\sum_{k=1}^{|R|}(-t)^k
\sum_{\{\mu_1,\ldots,\mu_k\}}e^{\mu_1+\cdots+\mu_k}
=\prod_{\mu\in\Lambda_R}(1-te^\mu)\,,
\end{align}
which of course is just the product of the characters for the
individual fermions making up the representation $R$.
It then follows that
\begin{align}
\chi^B_R(t)\equiv\chi(\ZZ^B_R(t))=(\chi^F_R(t))^{-1}
=\prod_{\mu\in\Lambda_R}(1-te^\mu)^{-1}\,.
\end{align}
The character picture will be used for a proof of our result in
Section \ref{Denominator}.

The examples above used for setting the notation are
valid only
for unconstrained variables (creation operators). 
We will use such refined partition functions to encode the spectrum of
generators in the Borcherds superalgebras described
in Section~\ref{SuperAlgebraSub}.

Before going into the construction of partition functions for algebras
and for constrained objects, we will consider two other situations, 
which will be of use later. The first is when a fermionic or bosonic
variable is ``maximally constrained'', so that any bilinear
vanishes. Then the partition function just contains a linear term:
\begin{align}\label{MaxConstrEq}
\ZZ(t)=1+\sigma Rt\,
\end{align}
(where again $\sigma=\pm1$ for bosons and fermions, respectively).
The second situation concerns variables of ``indefinite statistics'',
meaning that both symmetric and antisymmetric products of $R$
occur (but with odd levels still labeled as bosonic or fermionic by a sign $\sigma$).
Then the partition function is
\begin{align}\label{UnSpecStatEq}
\ZZ(t)=\bigoplus\limits_{p=0}^\infty(\sigma t)^p\text{\large{$\otimes$}}^pR
     =(1-\sigma Rt)^{-1}\,.
\end{align}
The observation that the partition functions (\ref{MaxConstrEq}) and (\ref{UnSpecStatEq}) 
are each other's inverses for opposite choices of $\sigma$ is one,
somewhat trivial, example of our main result which will be demonstrated in the
following sections. In this case the algebra is freely generated by
the representation at level one.

%%%%%%%%%%%%%%%%%%%%%%%%%%%%%%%%%%%%%%%%%%%%%%%%%%%%%%%%%%%%%%%%%%%%%%%%%%%%

\section{BRST operator and coalgebra}\label{brst-section}

%%%%%%%%%%%%%%%%%%%%%%%%%%%%%%%%%%%%%%%%%%%%%%%%%%%%%%%%%%%%%%%%%%%%%%%%%%%%

Consider the 
subalgebra $\BB_+$ of elements at positive levels of a ${\mathbb
  Z}$-graded Borcherds superalgebra $\BB$, as defined in Section
\ref{SuperAlgebraSub}. 
In the generic case, the algebra will be infinite-dimensional, and
contain elements at arbitrarily high levels. However, as we saw in Section \ref{SuperAlgebraSub}, all this
information is contained in the covariant Serre relations
\begin{align}\label{CovSerreEq}
\{E_M,E_N\}|_{R_2^\perp}=0\,,
\end{align}
where $R_2{}^\perp$ is the complement to $R_2$ in $\vee^2R_1$.
At level two, we thus have
generators $E_{MN}=\{E_M,E_N\}$ in $R_2$.

As announced in Section \ref{Introduction}, 
we will argue that all information about the
representations occurring at each level can be obtained in an alternative way,
which often provides a more direct answer, namely by considering a bosonic
object $\lb_M$ in $R_1$, subject to the constraint
\begin{align}
\lb^2|_{R_2}=0\,.
\end{align}
Notice that the object $\lb_M$ has opposite statistics (bosonic) to
$E_M$ (thinking of odd elements in a superalgebra as fermionic),
and that its constraint is in the symmetric 
representation complementary to that of the Serre relations.
The precise relation we will establish, and which is the main result
of this paper, is:
\\
\vskip\parskip
\boxit{
\vtop{\narrower{\null\vskip3\parskip 
\noindent\it The partition function of the universal enveloping
  algebra $U(\BB_+)$ is the inverse of the 
partition function for the
  constrained object $\lb$, \ie,\smallskip}}
\begin{align}\label{MainResult}
\ZZ_{U(\BB_+)}(t)\otimes\ZZ_\lb(t)=1\,.
\end{align}
\vskip-\parskip
}

\noindent Since the partition functions used are completely refined, in the
sense of Section \ref{PartFuncSub}, this provides complete information
of the generators at each level of the Borcherds superalgebra $\BB$.
The refined partition function for $\lambda$, if $\lambda$ is seen as
a complex object, can be seen as encoding holomorphic functions of $\lambda$.

The way we will argue for this equality in the present section 
is by identifying the action of the BRST operator for the
(conjugated)
constraint with the operation ``$d$'' of the coalgebra $\BB_+^*$.
This will not constitute a full proof (which would require a
consideration of cohomology of $\BB_+^*$), but provides a clear picture
of the correspondence. The proof, based on the denominator formula for
Borcherds superalgebras, is given in Section \ref{Denominator}.

Let us first consider the coalgebra, repeat some well known facts and
set the notation. For simplicity, 
we do this for the case of an ordinary Lie algebra;
the generalisation to graded brackets and Lie superalgebras is
trivial.
The coalgebra of a Lie algebra $\af$ is defined on the
vector space $\af^*$ dual to $\af$. 
It is equipped with a map
$d:\,\af^\ast\rightarrow\af^\ast\wedge\af^\ast$,   
which is dual to the Lie bracket $[\cdot,\cdot]$ in the sense that for any 
$A,B\in\af$ and $X\in\af^*$,
\begin{align} \label{*}
\la \,dX\,|\,A\wedge B \,\ra\,=\,\la \,X\,|\,[A,B]\,\ra\,,
\end{align}
where $\la\cdot|\cdot\ra$ is the canonical scalar product, naturally
extended to tensor products.
If $E_a$ and $E^{*a}$ are dual bases for $\af$ and $\af^*$, and 
$[E_a,E_b]=f_{ab}{}^cE_c$, eq. (\ref{*}) reads
\begin{align}
dE^{*a}=f_{bc}{}^a E^{*b}\wedge E^{*c}\,.
\end{align}
The action of $d$ is naturally extended to tensor products of elements
by defining it to act as a derivation.
The Jacobi identity is equivalent to the nilpotency, $d^2=0$, of $d$.
The above can be generalised to a Lie
superalgebra with the appropriate graded interpretation of wedge
products, brackets and derivations.

We now specialise on the Borcherds superalgebras at hand.
The first two levels of the coalgebra $\BB_+^*$ read
\begin{align}
dE^{*M}&=0\,,\cr
dE^{*MN}&=E^{*M}\vee E^{*N}|_{\bar R_2}\,.\label{**}
\end{align}
The Serre relations manifest themselves as the absence of 
generators in $\bar R_2{}^\perp$ at level two. 
What is the procedure for the continued
construction? Of course, knowledge of the algebra directly provides the
full information of the coalgebra. But it is also possible
to use eq. (\ref{**}) as a starting point for recursively deriving the
content at each level as well as the coproduct. One must then allow
for the most general representation for $E^{*(3)}$ and the most general form
of $dE^{*(3)}\sim E^{*(2)}\wedge E^{*(1)}$ consistent with
$d^2=0$. 
A general Ansatz consists of letting $E^{*MNP}$ belong to a
representation $R_3\subset R_1\otimes R_2$ and writing
\begin{align}
dE^{*MNP}=E^{*M}\wedge E^{*NP}|_{\bar R_3}\,.
\end{align}
The nilpotency of $d$ then determines the allowed representation $R_3$. For
example, a totally symmetric representation is always excluded from $R_3$,
since it will vanish due to the Jacobi identity.
This procedure can then be continued to all levels, where $dE^{*(p)}$
will contain sums of terms $E^{*(q)}\wedge E^{*(p-q)}$ (wedge here
denoting graded antisymmetrisation).

The unique result of the procedure can be understood by the following
argument, which also provides a conceptual idea behind the result stated in
eq. 
(\ref{MainResult}).
Everything starts from, and is generated from, the basic set of
generators $E_M$ in the $\gf$-representation $\BB_1$. 
Since they are odd elements of a superalgebra,
they are normally thought of as fermionic. It is however useful to
think of $E_M$ as not carrying a definite statistics. Indeed,
considering the Serre relations (\ref{CovSerreEq}), 
the only constraint on a
bilinear in $E_M$ (including both symmetric and anti-symmetric
parts) is that a certain representation $R_2{}^\perp$ of the symmetric bilinear
vanishes. The identification of the symmetric part in the complement $R_2$ with
``new'' generators $E_{MN}$
is in this sense optional.
Without this identification, and corresponding identifications at higher levels,
the universal algebra $U(\BB_+)$ can be constructed from the tensor
algebra of $\BB_1$ by factoring out the ideal 
generated by $E_M \otimes E_N |_{R_2{}^\perp}$.
This provides
a way of constructing an arbitrary element, not in the algebra
$\BB_+$, but in its universal enveloping algebra $U(\BB_+)$, in terms
of powers of $E_M$ only.
The partition function of the universal enveloping algebra 
will be that of an object $E_M$ in $R_1$ {\it of indefinite
  statistics} (although the elements
at odd levels are labelled as fermionic in the partition function,
see Section \ref{PartFuncSub}),
 modulo the ideal
generated by the Serre relations. Seen this way, our main result can
be phrased in the following way:

{\it The partition function for a bosonic object ($\lambda$) 
in $R_1$ subject to a
bilinear constraint in $R_2$ is the inverse of the partition function
for an object (the set of level-one generators in $\BB$) 
with indefinite statistics, where odd powers are labeled
as fermionic, subject to a bilinear constraint in $R_2{}^\perp$.}

This statement provides an interpolating generalisation for
partitions of constrained objects of the ones made for unconstrained
and maximally constrained ones in Section
\ref{PartFuncSub}. However, unlike in those limiting cases, the
statistics here may not be switched, which we will comment on in
Section \ref{WhyNotLie}.

Now, consider an object $\bar\lb$ in $\bar R_1$, with the constraint 
$\bar\lb^2|_{\bar R_2}=0$. The constraint can be treated using a BRST
formalism. For convenience, we change our notation and use $c^M$
instead of $\bar\lb^M$. The first term in the BRST operator $Q$ is
$Q_{(2)}=b_{MN}c^Mc^N$, where $b_{MN}$ in $R_2$ is
the ghost for the constraint\footnote{It would maybe be more
conventional to use a notation where $c$ is the ghost multiplying the
constraint, and $b$ its conjugate. Here, however, it turns out that
all terms will be of the form $bcc$, which corresponds to the standard
form of ``algebra'' ghost terms in a BRST operator.}. 
However, if the constraint happens to be
reducible, there will be higher order ghosts compensating for the
reducibility.
Such reducibility will be captured by the introduction of a new $bc$ pair,
and a term
$Q_{(3)}=b_{MNP}c^{MN}c^P$
in $Q$. The representation of $b_{MNP}$ is everything that is allowed by
$Q^2=0$. This should be continued, as long as the reducibility
continues, \ie, as long as further such terms can be added.
A generic term will be of the form
$b_{(p+q)}c^{(p)}c^{(q)}$, where 
the ghosts are alternatingly fermionic and bosonic.
From this trilinear form of the BRST operator it is immediately clear
that its action on the $c$ ghosts defines the coalgebra of a Lie
superalgebra. An infinite reducibility\footnote{The concept of
  reducibility is not absolute, but may depend on the degree of
  covariance. Here, we always consider reducibility as expressed in
  terms of representations of the
  finite-dimensional Lie algebra $\gf$ (but should of course not be
  confused with the possible reducibility of the representations
  themselves).} corresponds to an 
infinite-dimensional algebra.

We now recognise the exact parallel between on one hand the construction of the
coalgebra, given the Serre relations (and nothing more), and on the
other hand
the construction of the BRST operator. The difference is only a matter of
notation. The cogenerators $E^{*(p)}$
correspond to the ghosts $c^{(p)}$, and the graded wedge products are
automatically implied by the 
``wrong'' statistics of the ghosts. The operator $d$ is the adjoint action
(graded commutator) of $Q$, so that $dE^{*(p)}\leftrightarrow[Q,c^{(p)}\}$.

This means that if we calculate the partition function of
$\lambda$ 
as a constrained object, which is obtained as the conjugate of the tensor
product of the partition functions of all the ghosts, 
\begin{align}
\ZZ_\lambda(t)=\bigotimes\limits_{p=1}^\infty
     (1-t^p)^{(-1)^pR_p}\,,
\end{align}
it will coincide
with the inverse partition function of the universal enveloping
algebra $U(\BB_+)$, which by definition is
\begin{align} \label{R-partfn}
\ZZ_{U(\BB_+)}(t)=\bigotimes\limits_{p=1}^\infty
     (1-t^p)^{(-1)^{p+1}R_p}\,,
\end{align}
using the shorthand notation of (\ref{ZZRFDef}).
The inverse simply appears since the
correspondence $E^{*(p)}\leftrightarrow c^{(p)}$ changes statistics.

The above argument does not provide a strict proof of
eq. (\ref{MainResult}). The missing step is the proof that the BRST
operator $Q\sim bcc$ correctly encodes the degrees of
freedom of the constrained object, or, equivalently, that no other
unwanted cohomology arises. We refrain from doing this, but we will
present a different proof in Section \ref{Denominator}. 

Neither of the two above methods of finding the spectrum of generators
has an advantage over the other, since we just demonstrated that they
contain exactly the same calculational steps. However, knowing that
the partition function is that of a constrained object $\lb$ can often
provide an alternative, more direct, and simpler way of obtaining the
answer. 
Provided that we know from the constraint which representation $S_p$ appear
at any power $\lb^p$, the partition function is directly constructed as
\begin{align} \label{S-partfn}
\ZZ_{\lb}(t)=\bigoplus\limits_{p=0}^\infty S_p t^p\,. 
\end{align}
Expanding this partition function in a product 
form\footnote{This is the method used by Berkovits and Nekrasov in
  ref. \cite{Berkovits:2005hy} to obtain detailed information on the
  partition functions 
of pure spinors.}
gives information
about all the ghost representations, and thus about the generators of the
algebra.
This calculation becomes especially simple in cases where $S_1=R_1$ is an
irreducible representation of some Lie algebra with highest weight $\lambda$, 
and $S_2=R_2{}^\perp=\vee^2R_1\ominus R_2$ 
is the representation with highest weight $2\lambda$. 
Then $S_p$ will have highest weight $p\lambda$.
Indeed, the class of Borcherds superalgebras we consider all have this
property, as will be shown in Section \ref{Denominator}.
We will give some examples of such situations
in Section \ref{ExamplesSect}, among which are pure spinors and their
associated superalgebras.
Although the representations $R_p$ are complicated, they can be calculated
from the more readily available representations $S_p$
by inserting eqs. (\ref{R-partfn})--(\ref{S-partfn}) into
(\ref{MainResult}), which gives 
\begin{align} \label{RS}
\bigotimes\limits_{p=1}^\infty(1-t^p)^{(-1)^pR_p}
=\bigoplus\limits_{q=0}^\infty S_q t^q\,.
\end{align}
The explicit solution of this relation for the spectrum of the
Borcherds superalgebra, \ie, the representations
$R_p$ in terms of the known $S_p$,
can be obtained by recursion, or by a M\"obius inversion.
By comparing the left and right hand sides for the first few powers of
$t$ we get 
\begin{align} \label{recformulas}
R_1&=S_1\,,\nn\\
R_2&= \text{\large{$\vee$}}^2 R_1\ominus S_2\,,\nn\\
R_3&=(R_1 \otimes R_2)\ominus\text{\large{$\vee$}}^3 R_1\oplus S_3\,,\nn\\
R_4&= \big((R_1 \otimes R_3)\oplus \text{\large{$\wedge$}}^2 R_2 \big) \ominus (\text{\large{$\vee$}}^2 R_1\otimes R_2) \oplus 
\text{\large{$\vee$}}^4 R_1 \ominus S_4\,.
\end{align}
We will display some explicit examples of varying complexity in
Section \ref{ExamplesSect}.

%%%%%%%%%%%%%%%%%%%%%%%%%%%%%%%%%%%%%%%%%%%%%%%%%%%%%%%%%%%%%%%%%%%%%%%%%%%%

\section{A proof from denominator formulas}\label{Denominator}

%%%%%%%%%%%%%%%%%%%%%%%%%%%%%%%%%%%%%%%%%%%%%%%%%%%%%%%%%%%%%%%%%%%%%%%%%%%%

This section will provide a proof of our main result
(\ref{MainResult}), using the
denominator formula for Borcherds
superalgebras \cite{Ray95,Miyamoto,Ray,Wakimoto}.
It is known for general Borcherds
superalgebras but here we only need a simplified version given below, valid for the special cases
of Borcherds
superalgebras under consideration.

Let $\Phi$ be the root system of $\BB$, and 
for any integer $p$, let $\Phi_p$ be the subset of $\Phi$
consisting of all roots 
$\beta=p\beta_0+\alpha$, where $\alpha$ is a linear combination of the
real simple roots $\beta_i=\alpha_i$. 
Thus $\Phi_0$ is the root system of the subalgebra $\g$,
and $\BB_p$ is the direct sum of all root spaces $\BB_\beta$ such that
$\beta \in \Phi_p$, and, if 
$p=0$, the Cartan subalgebra.

We will show that the eq. (\ref{RS}), with the
lowest weights of the representations $S_q$ given by the Dynkin labels 
\begin{align} \label{dynkinlabels}
\lambda_i=q(\beta_0,\beta_i{}^\vee)=q \cdot
2\frac{(\beta_0,\beta_i)}{(\beta_i,\beta_i)}=qB_{i0}\,, 
\end{align}
is equivalent to the denominator formula for $\BB$ \cite{Ray95,Miyamoto,Ray,Wakimoto}, which reads
\begin{align} \label{denominator-formula}
\frac{\prod_{\beta \in \Phi_{(0)}^+} (1-e^{-\beta})^{\rm mult\,
    \beta}}{\prod_{\beta \in \Phi^+_{(1)}} (1+e^{-\beta})^{\rm mult\,
    \beta}} 
&= \sum_{w \in W} \sum_{q=0}^\infty (-1)^{|w|}(-1)^q e^{w(\rho-q\beta_0)-\rho}\,.
\end{align}
Here $\Phi^+_{(0)}$ and $\Phi^+_{(1)}$ consist of all even and odd
positive roots, respectively, 
$\rho$ is the Weyl vector of $\cB$, defined by 
\begin{align}
(\rho,\beta_I)=\frac{(\beta_I,\beta_I)}{2},
\end{align}
and the Weyl group $W$ of $\BB$ is generated by all fundamental Weyl reflections
\begin{align}
r_i: \beta \mapsto \beta-(\alpha_i{}^\vee,\beta)
=\beta-2\frac{(\alpha_i,\beta)}{(\alpha_i,\alpha_i)}\,.
\end{align}
The length $|w|$ of an element $w$ in $W$ is the minimal number of
fundamental Weyl reflections (not necessarily distinct), which, 
applied after each other, give $w$.

Note that
the representations $S_p$ are given by the Dynkin labels $\lambda_i$
of their {\it lowest} weights, since we consider positive 
levels of $\BB$. However, we are going to relate the denominator
formula (\ref{denominator-formula}) for $\BB$ to the character formula
for $\g$, 
which is usually expressed in terms
of the highest {\it weight} of a representation. Therefore it is convenient
to replace eq. (\ref{RS}) 
by the equivalent equation obtained by conjugating all
representations,
\begin{align} \label{RS-conj}
\bigotimes\limits_{p=1}^\infty(1-t^p)^{(-1)^p\bar{R}_p}
=\bigoplus\limits_{q=0}^\infty \bar{S}_q t^q\,,
\end{align}
where now the highest weight of $\bar{S}_q$ is
$-\lambda_i=-qB_{i0}$. What we will actually show is that this 
eq. (\ref{RS-conj}) is equivalent to the denominator formula
(\ref{denominator-formula}).\footnote{Instead of considering positive
  levels in $\BB$ and conjugating the representations we could of
  course also have considered 
negative levels and only highest weights from the beginning.}

Let $\tilde{\Lambda}{}_0$ be the element in the weight space of
$\BB$ such that $(\tilde{\Lambda}{}_0,\alpha_i)=0$ for all
$i=1,\ldots,r$, and the componenent of $\tilde{\Lambda}{}_0$
corresponding to $\beta_0$ in the basis of
simple roots is 
equal to one (this element exists uniquely since both $B_{IJ}$ and $B_{ij}=a_{ij}$ are non-degenerate).
Thus $\tilde{\Lambda}{}_0-\beta_0$ is an element in the
weight space of $\g$ (considered as a subspace of the weight space of
$\BB$). 
More generally, a root $\beta \in \Phi_p$ can be written
$\beta=p\tilde{\Lambda}{}_0+\mu$, where
$\mu=\beta-p\tilde{\Lambda}{}_0$ 
is an element in the weight space of $\g$. We then get
\begin{align}
\prod_{\beta \in \Phi_p} (1-e^{-\beta})^{\rm mult\, \beta} &=
\prod_{\mu \in R_p} (1-e^{-p\tilde{\Lambda}{}_0}e^{-\mu})^{\rm mult\,
  \mu}\nn\\ 
&= \sum_{k=0}^{|R_p|} (-1)^k\sum
e^{-(\mu_1+\cdots+\mu_k)}(e^{-\tilde{\Lambda}{}_0})^{kp}\,, 
\end{align}
where the second sum goes over all sets of $k$ distinct weights
$\mu_1,\ldots,\mu_k$
among the weights of $R_p$,  counting (as in Section
\ref{PartFuncSub}) a weight with multiplicity $m$ as 
$m$ ``distinct'' weights.
This sum can be obtained from the character for $\bwedge^kR_p$ by
inverting each term, which corresponds to conjugating the
representation 
$R_p$.  Thus
\begin{align}
\prod_{\beta \in \Phi_p} (1-e^{-\beta})^{\rm mult\, \beta} &=
\sum_{k=0}^{|\bar R_p|} (-1)^k \chi (\bwedge^k\bar R_p)
s^{kp} 
\end{align}
where we have set $s=e^{-\tilde{\Lambda}{}_0}$. In the same way,
\begin{align}
\prod_{\beta \in \Phi_p} (1+e^{-\beta})^{\rm mult\, \beta} &=
\sum_{k=0}^{|\bar R_p|} \chi (\bwedge^k\bar R_p) s^{kp} 
\end{align}
and we know that the inverse of this is
\begin{align}
\prod_{\beta \in \Phi_p} (1+e^{-\beta})^{-\rm mult\, \beta} &=
\sum_{k=0}^{\infty} (-1)^k \chi (\bvee^k\bar R_p) s^{kp}\,. 
\end{align}
Here the character of $\vee^k \bar R_p$ is given by the sum of all terms 
$e^{-(\mu_1+\cdots+\mu_k)}$, where $\mu_1,\ldots,\mu_k$
are weights of $R_p$, this time 
not necessarily distinct.
Following the notation in Section \ref{PartFuncSub}, we write this as
\begin{align}
\prod_{\beta \in \Phi_p} (1\pm e^{-\beta})^{\mp\rm mult\, \beta} &=
\chi\big((1\pm s^p)^{\mp \bar R_p}\big),
\end{align}
and the left hand side of eq. (\ref{denominator-formula}) becomes
\begin{align} \label{finallhs}
\frac{\prod_{\beta \in \Phi_{(0)}^+} (1-e^{-\beta})^{\rm mult\,
    \beta}}{\prod_{\beta \in \Phi^+_{(1)}} (1+e^{-\beta})^{\rm mult\,
    \beta}} 
&= \prod_{\alpha \in \Phi_{0}^+} (1-e^{-\alpha})^{\rm mult\,
  \alpha}\prod_{p=1}^\infty 
\chi\Big(\big(1-(-1)^ps^p\big)^{(-1)^p \bar R_p}\Big)\,,
\end{align}
where $\Phi_{0}^+$ consists of the positive roots of $\g$.

We now turn to the right hand side of the denominator formula,
\begin{align} \label{rhs}
\sum_{w \in W} \sum_{q=0}^\infty (-1)^{|w|}(-1)^q e^{w(\rho-q\beta_0)-\rho}.
\end{align}
Here $W$ is the Weyl group of $\BB$, but since it is generated by the
fundamental Weyl reflections corresponding to the real roots only, it
coincides with 
the Weyl group of $\g$. 
In order to use the character formula for $\g$ we also need to replace
the Weyl vector of $\BB$ with the one of $\g$, but this requires some more consideration. 
The Weyl vector $\rho=\rho_\BB$ of $\BB$ is defined as the element
in the weight space of $\BB$ satisfying
\begin{align}
(\rho_\BB,\beta_I)=\frac{(\beta_I,\beta_I)}{2}\,,
\end{align}
whereas the Weyl vector $\rho_\g$ of $\g$ only has to satisfy
\begin{align}
(\rho_\g,\alpha_i)=\frac{(\alpha_i,\alpha_i)}{2}\,,
\end{align}
but on the other hand it must have a zero component corresponding to
$\beta_0$ in the basis of simple roots. 
Thus the Weyl vectors of $\BB$ and $\g$ are different (in general),
but since their difference $\rho_\BB-\rho_\g$ is orthogonal to the
real roots, 
$(\rho_\BB-\rho_\g,\alpha_i)=0$, it is invariant under the Weyl group,
$w(\rho_\BB-\rho_\g)=\rho_\BB-\rho_\g$. 
We then get
\begin{align}
w(\rho_\BB-q\beta_0)-\rho_\BB&=w\big(\rho_\g+(\rho_\BB-\rho_\g)
-q\beta_0\big)-\rho_\g-(\rho_\BB-\rho_\g)\nn\\&=w(\rho_\g-q\beta_0)-\rho_g  
+w(\rho_\BB-\rho_\g)-(\rho_\BB-\rho_\g)\nn\\&=w(\rho_\g-q\beta_0)-\rho_g
\end{align}
and we can indeed replace $\rho=\rho_\BB$ by $\rho_\g$ in eq.
(\ref{rhs}). To simplify the notation, we will henceforth write
$\rho=\rho_\g$. 
Furthermore, since also $\tilde{\Lambda}{}_0$ is orthogonal to the
real roots, we have 
\begin{align}
w(\rho-q\beta_0)-\rho&=
w\big(\rho-q\tilde{\Lambda}{}_0+q(\tilde{\Lambda}{}_0-\beta_0)\big)-\rho\nn\\ 
&= w\big(\rho+q(\tilde{\Lambda}{}_0-\beta_0)\big)-\rho-q\tilde{\Lambda}{}_0
\end{align}
and then
\begin{align} \label{finalrhs}
\sum_{w \in W} \sum_{q=0}^\infty (-1)^{|w|}(-1)^q e^{w(\rho-q\beta_0)-\rho}
&= \sum_{w \in W} \sum_{q=0}^\infty (-1)^{|w|}(-1)^q
e^{w(\rho+q(\tilde{\Lambda}{}_0-\beta_0))-\rho-q\tilde{\Lambda}{}_0}\nn\\ 
&= \sum_{w \in W} \sum_{q=0}^\infty (-1)^{|w|}(-1)^q
e^{w(\rho+q(\tilde{\Lambda}{}_0-\beta_0))-\rho}s^q\,. 
\end{align}
Equating eqs. (\ref{finallhs}) and (\ref{finalrhs}) we get
\begin{align}
\prod_{p=1}^\infty \chi\Big(\big(1-(-1)^ps^p\big)^{(-1)^p\bar R_p}\Big)=
\sum_{q=0}^\infty\frac{\sum_{w \in
    W}(-1)^{|w|}e^{w(\rho+q(\tilde{\Lambda}{}_0-\beta_0))-\rho}} 
{\prod_{\beta \in \Phi_{0}^+}(1-e^{-\beta})^{\rm mult\, \beta}}(-1)^qs^q\,,
\end{align}\,
where we recognise
\begin{align}
\frac{\sum_{w \in W}(-1)^{|w|}e^{w(\rho+q(\tilde{\Lambda}{}_0-\beta_0))-\rho}}
{\prod_{\beta \in \Phi_{0}^+}(1-e^{-\beta})^{\rm mult\, \beta}}
\end{align}
as the character of the representation 
of $\g$ with highest weight $q(\tilde{\Lambda}{}_0-\beta_0)$ given by the Dynkin labels
\begin{align}
\big(q(\tilde{\Lambda}{}_0-\beta_0),\beta_i{}^\vee\big)=-q(\beta_0,\beta_i{}^\vee) 
=-q \cdot 2\frac{(\beta_0,\beta_i)}{(\beta_i,\beta_i)}=-qB_{i0}=-\lambda_i\,,
\end{align}
and thus
\begin{align}
\prod_{p=1}^\infty \chi\Big(\big(1-(-1)^ps^p\big)^{(-1)^p\bar R_p}\Big)=
\sum_{q=0}^\infty \chi(\bar S_q)
(-1)^qs^q.
\end{align}
Finally, substituting $s$ by $-t$ we arrive at the equation
\begin{align}
\prod_{p=1}^\infty \chi\big((1-t^p)^{(-1)^p\bar R_p}\big)=
\sum_{q=0}^\infty
\chi(\bar S_q) t^q\,,
\end{align}
which is the character version of (and thus equivalent to) eq. (\ref{RS-conj}).

%%%%%%%%%%%%%%%%%%%%%%%%%%%%%%%%%%%%%%%%%%%%%%%%%%%%%%%%%%%%%%%%%%%%%%%%%%%

\section{Why is the method not applicable to Lie algebras?}\label{WhyNotLie}

%%%%%%%%%%%%%%%%%%%%%%%%%%%%%%%%%%%%%%%%%%%%%%%%%%%%%%%%%%%%%%%%%%%%%%%%%%%

Let us replace $\beta_0$ with an even simple root $\alpha_0$, which is real, $(\alpha_0,\alpha_0)>0$,
but otherwise satisfies the same inner product relations as $\beta_0$, thus $(\alpha_0,\alpha_i)=(\beta_0,\alpha_i)$.
The Chevalley-Serre relations (\ref{chev-rel-1})--(\ref{serre-rel1}),
with all superbrackets being ordinary antisymmetric brackets, and with the Cartan matrix $B_{IJ}$ replaced by
\begin{align}
A_{IJ}=(\alpha_J,\alpha_I{}^\vee)
=2\frac{(\alpha_J,\alpha_I)}{(\alpha_I,\alpha_I)},
\end{align}
defines a Kac-Moody algebra $\mathcal A$.
This corresponds to adding an
ordinary (white) node to the Dynkin diagram of $\mathfrak g$ instead of a grey one, and
in analogy with $\BB$ the adjoint representation of $\mathcal A$ dcomposes
into $\mathfrak g$-representations $R_p$. The representation $R_1$ is the same
as in the case of
$\cB$, but $R_2$ is now a subrepresentation of $\wedge^2R_1$, the {\it anti-symmetric} tensor product of $R_1$ with itself.
Its complement is the direct sum of representations with lowest weights given by the Dynkin labels
$\lambda_i=A_{ij}+2A_{i0}$
for all $j$ such that $A_{0j}\neq0$.

One might imagine that the statement (\ref{MainResult}) would apply
both for the Lie superalgebra $\BB$ and the ordinary Lie algebra $\mathcal A$. This
would potentially have made it possible to extract precise information
about the generators to all levels for classes of infinite-dimensional
(\eg\
hyperbolic) Kac--Moody algebras. It turns out, however, that statistics
can not just be changed. This is because the constrained object
$\lambda$ then would be fermionic. Having bilinear (bosonic)
constraints on fermionic variables is generically a strange situation,
and leads to complicated structures,
as we will explain.

Consider a fermionic $\lambda$ in a representation $R_1$ of $\gf$ with
a bilinear 
constraint in $R_2$ of $\mathcal A$, thus complementary to some Serre relations in the 
  anti-symmetric product $\wedge^2R_1$.
An algebraic ``solution'' to the constraints (in the sense that one
considers a power series in $\lambda$ modulo the constraint) 
will result in a polynomial
partition function, where the highest term is of order lower than or equal to
$|R_1|$.
Its factorisation in ghost contributions is however infinite. This is
because the ghosts, like the original variables, are fermionic, and so
are the ghosts for reducibility. Instead of removing fermionic degrees
of freedom, the ghosts add more fermions, corresponding to the removal
of bosonic degrees of freedom (the constraint). 

This somewhat pathological 
behaviour is in itself not an obstruction for the existence 
of a relation like eq. (\ref{MainResult}) --- one might well imagine
that a properly regularised sum with strictly positive terms yields a
negative value (although it {\it is} a valid argument against an
analogous construction when the Lie algebra is finite-dimensional). 
What makes things go wrong
is the fact that a bilinear constraint on fermions inherently has some
reducibility coming from the fermionic property of the
variables. Whatever the bosonic constraint is, it is \eg\ obvious that
raising it to a sufficient power will give zero due to saturation of
fermions. This has no counterpart in the bosonic situation, and will
introduce ghosts in a BRST treatment which do not enter $Q$ in the ``$bcc$''
form. Therefore, a correct BRST treatment can not be given an
interpretation in terms of a Lie algebra. 
We have observed in a number of examples 
that a naive treatment of the Serre relations as complementary to a
constraint
gives agreement in the spectrum to a number of low levels, 
before the ``saturation of fermions'' becomes relevant.
Whether there is a systematic way of consistently defining partitions
for fermions that circumvents this and correctly encodes the Serre
relations (and thereby the spectrum of the Lie algebra) is an open
question. In any case it seems reasonable that the occurrence, in the
superalgebra case, of the
highest weights which are simply multiples of the defining one, has
particularly simple structure without counterpart in the Lie algebra
situation.

Turning to the actual proof of the main result for the Lie superalgebra $\BB$ in
Section \ref{Denominator}, it is easy
to identify the step where the argument fails for Lie algebras. The
Weyl group of the extended algebra $\cA$ is not identical to that of
$\g$, and the proof in its present form fails, 
although the denominator formula is known.

%%%%%%%%%%%%%%%%%%%%%%%%%%%%%%%%%%%%%%%%%%%%%%%%%%%%%%%%%%%%%%%%%%%%%%%%%%%

\section{Examples}\label{ExamplesSect}

%%%%%%%%%%%%%%%%%%%%%%%%%%%%%%%%%%%%%%%%%%%%%%%%%%%%%%%%%%%%%%%%%%%%%%%%%%%

\renewcommand{\cal}{\mathscr}

We will give a number of examples that illustrate the connection
between the constrained bosonic variable and the spectrum of
generators in the superalgebra. 

We use the notation $\BB_+$ for the subalgebra of generators at
positive levels, although in some examples (the freely generated
algebras) it is not a subalgebra of a Borcherds
superalgebra of the precise type described in Section
\ref{SuperAlgebraSub}.

\subsection{The extreme cases}\label{ExtremeSub}
Consider first a freely generated superalgebra. Then the Serre relations
are empty, and elements of the universal enveloping algebra $U(\BB_+)$ 
are given
by arbitrary tensor products of $R_1$. Thus $\ZZ_{U(\BB_+)}$ is given by
eq. (\ref{UnSpecStatEq}) with $\sigma=-1$, which is the inverse of the
formal partition for a ``maximally constrained'' boson in $R_1$. This
is one extreme case of the correspondence (in which the statistics can
be interchanged). It appears if the additional simple root is not a null root, but has {\it negative} length squared.

The other extremal case is when the Serre relations fill the whole
symmetric product $\vee^2R_1$, so that
$\{E_M,E_N\}=0$. The superalgebra is then
finite-dimensional with $E_M$ forming a basis for $\BB_+=\BB_1$. The
partition function $\ZZ_{U(\BB_+)}$ is the partition function of
fermions in $R_1$, which is the inverse of the partition function for
an unconstrained boson. Also in this other special case the statistics
can be interchanged.

Intermediary cases only work as a correspondence of the form
(\ref{MainResult}) between superalgebras and constrained bosons, and
these provide less trivial illustrations of our result.

\subsection{{\it D} {\rm = 8} pure spinors and null vectors}\label{NullVectorsSub}
Pure spinors provide well known and extensively studied 
examples of constrained bosons.
They lead to
minimal spinor orbits under Spin
groups, due to the fact that only a single irreducible representation
appears in a spinor
bilinear (the one whose 
highest weight is twice the one of a spinor), and by induction there is only one representation for each positive
power of the spinor. We will give two examples of pure spinors, in this subsection and the next.

Let us first consider a pure spinor in $D=8$, where the constraint is
particularly simple, $\lambda_\alpha\lambda_\alpha=0$. This is via
triality equivalent to a null vector.
The Dynkin diagram of the corresponding superalgebra is given below.
\noindent
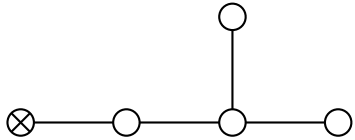
\begin{center}
\begin{picture}(130,50)(55,-5)
\thicklines
\put(60,0){\circle{10}}
\put(56.5,-3.5){\line(1,1){7}}
\put(56.5,3.5){\line(1,-1){7}}
\put(100,0){\circle{10}}
\put(140,0){\circle{10}}
\put(180,0){\circle{10}}
\put(140,40){\circle{10}}
\put(65,0){\line(1,0){30}}
\put(105,0){\line(1,0){30}}
\put(145,0){\line(1,0){30}}
\put(140,5){\line(0,1){30}}
\end{picture} 
\end{center}
The analysis can equally well be performed for null vectors in general
dimension $D$.
The refined partition function for $\lambda$ reads
\begin{equation}
\ZZ_\lambda(t)=\bigoplus_{p=0}^\infty(p0\ldots0)t^p\,,
\end{equation}
where Dynkin labels of highest weights have been used for the
representations, and $(10\ldots0)$ denotes the vector representation. 
The representation 
$(p0\ldots0)$ consists of symmetric and traceless multi-vectors. Its
dimension is readily calculated to be 
\begin{equation}
\binom{p+D-1}{p}-\binom{p+D-3}{p-2}=\frac{(2p+D-2)(p+D-3)!}{(D-2)!p!}\,,
\end{equation}
so the unrefined partition function (just counting dimensions) is
\begin{equation}
Z_\lambda(t)=\sum_{p=0}^\infty\frac{(2p+D-2)(p+D-3)!}{(D-2)!p!}t^p
=\frac{1+t}{(1-t)^{D-1}}=\frac{1-t^2}{(1-t)^D}\,.
\end{equation}
The same result is obtained by constructing the partition function
from the ghosts. An (unconstrained) variable $\lambda$ contributes to
$\ZZ$ with $(1-t)^{-(10\ldots0)}$, and the fermionic ghost for the
constraint with $(1-t^2)^{(00\ldots0)}$. The constraint is
irreducible, so there are no higher ghosts.
The correspondence (\ref{MainResult}) tells us that the spectrum of
the universal enveloping algebra $U(\BB_+)$ is given by
\begin{equation}
\ZZ_{U(\BB_+)}(t)=(\ZZ_\lambda(t))^{-1}=(1-t)^{(10\ldots0)}
\otimes(1-t^2)^{-(00\ldots0)}\,,
\end{equation}
corresponding to a fermionic generator in $(10\ldots0)$ at level one and
a bosonic one in $(00\ldots0)$ at level two.
The superalgebra $\BB$, which here comes with a 5-grading, 
is finite-dimensional, $\BB\approx{\mathfrak{osp}}(D|2)$. A
finite-dimensional superalgebra is obtained when the reducibility of
the constraint on $\lambda$ is finite.

\subsection{{\it D} {\rm = 10} pure spinors and supergravity forms}\label{PureSpinorSub}
Let us turn to the more interesting cases of infinite-dimensional
superalgebras, which are related to the spectrum of forms in supergravity, and thereby also to the
tensor hierarchies in gauged supergravity (see the discussion in Section~1).

Pure spinors in $D=10$ are relevant for the off-shell superfield
formulation of $D=10$ super-Yang--Mills theory (see
{\it e.g.} refs. 
\cite{Cederwall:2001bt,Berkovits:2001rb,Cederwall:2001dx,Cederwall:2013vba}). 
The partition function is described in some detail in ref. 
\cite{Berkovits:2005hy},
and is given by
\begin{align}
\ZZ_\lambda(t)&=\bigoplus_{p=0}^\infty(0000k)t^p\cr
&=\bigl[(00000)\ominus(10000)t^2\oplus(00001)t^3\cr
&\quad\ominus(00010)t^5\oplus(10000)t^6\ominus(00000)t^8\bigr]
\otimes(1-t)^{-(00001)}\,,\label{PureSpinorPartition}
\end{align}
or, just counting dimensions,
\begin{equation}
Z_\lambda(t)=\frac{1-10t^2+16t^3-16t^5+10t^6-t^8}{(1-t)^{16}}
=\frac{(1+t)(1+4t+t^2)}{(1-t)^{11}}\,.
\end{equation}
The power 11 of the pole at $t=1$ signals 11 degrees of freedom in a
pure spinor.
The Dynkin diagram of the corresponding superalgebra is given below.
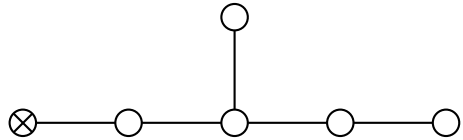
\begin{center}
\begin{picture}(170,50)(55,-5)
\thicklines
\put(60,0){\circle{10}}
\put(56.5,-3.5){\line(1,1){7}}
\put(56.5,3.5){\line(1,-1){7}}
\put(100,0){\circle{10}}
\put(140,0){\circle{10}}
\put(180,0){\circle{10}}
\put(220,0){\circle{10}}
\put(140,40){\circle{10}}
\put(65,0){\line(1,0){30}}
\put(105,0){\line(1,0){30}}
\put(145,0){\line(1,0){30}}
\put(185,0){\line(1,0){30}}
\put(140,5){\line(0,1){30}}
\end{picture} 
\end{center}
This algebra is infinite-dimensional. Still, we know that the spectrum is
determined by $\ZZ_{U(\BB_+)}=\ZZ_\lambda{}^{-1}$. The generators at
each level in $\BB_+$ are obtained by rewriting the
partition function (\ref{PureSpinorPartition}) on product form, which
reflects the ghost structure corresponding to the infinite
reducibility:
\begin{align}
(\ZZ_\lambda(t))^{-1}=\bigotimes\limits_{p=1}^\infty(1-t^p)^{(-1)^{p+1}R_p}\,.
\end{align}
This can be done recursively as in eq. (\ref{recformulas}), 
with the following result for the first few representations:
\begin{align}
&R_1=(00001)={\bf16},\,
R_2=(10000)={\bf10},\,
R_3=(00010)=\overline{\bf16},\,\cr
&R_4=(01000)={\bf45},\,
R_5=(10010)=\overline{\bf144},\,\cr
&R_6=(11000)\oplus(00020)\oplus(10000)
   ={\bf320}\oplus\overline{\bf126}\oplus{\bf10},\ldots
\end{align}
For the dimensionalities $|R_p|$, an explicit M\"obius inversion
formula can be found \cite{Berkovits:2005hy}.

The case of $D=10$ pure spinors is relevant to exceptional field
theory with
U-duality group $E_{5}\approx Spin(5,5)$.
Generally, the infinite ghost tower in exceptional field
theory with
U-duality group $E_{n}$ ($n\leq8$) is identical to the infinite spectrum of superforms in $D$-dimensional
maximal supergravity ($D=11-n$), as was shown for low levels in ref. \cite{Berman:2012vc}. Our results here, combined with those in
refs. \cite{Greitz:2011da,Greitz:2012vp,Howe:2015hpa}, establish this correspondence for all levels.
Here we have shown that the ghosts 
for a constrained object
give rise to a Borcherds superalgebra by the action of the BRST operator,
and in refs. \cite{Greitz:2011da,Greitz:2012vp,Howe:2015hpa}
it was shown that the forms in the supergravity theory similarly give rise to a Borcherds superalgebra by their Bianchi identities.
In the extended field theory the constraint is directly associated with the section
condition, and leads to the same Borcherds superalgebras as the supersymmetry constraint on the supergravity side.
Since the Borcherds superalgebras are the same, the sequences of representations are the same as well.

The (unrefined)
partition functions corresponding to the constraint in the exceptional field theories
were give in ref. \cite{Berman:2012vc}.
As an example, the $E_6$ case gives a Borcherds superalgebra
defined by a $\lambda$ belonging to a c\^one over the Cayley plane
\cite{Cederwall:1988hj}. The partition function is
\begin{align}
\ZZ_\lambda(t)&=\bigoplus_{k=0}^\infty(k00000)t^k\cr
&=\bigl[
(000000)\ominus(000001)t^2\oplus(010000)t^3\ominus(001000)t^5\cr
&\quad\oplus(100001)t^6\ominus(000002)t^7
\ominus(200000)t^8\oplus(100001)t^9\cr
&\quad\ominus(000010)t^{10}
\oplus(010000)t^{12}\ominus(100000)t^{13}\oplus(000000)t^{15}
\bigr]\,\cr
&\qquad\otimes(1-t)^{-(100000)}\,,\label{CayleyPlanePartition}
\end{align}
and the spectrum of the Borcherds algebra is obtained recursively by
rewriting $\ZZ_\lambda$ on product form.

With the same interpretation as gauge transformations and reducibility
for generalised diffeomorphisms, our earlier example with null vectors,
corresponding to a finite-dimensional superalgebra, is relevant for
double field theory with T-duality group $O(d,d)$.

\subsection{Superalgebras and Lie algebras}\label{LieSuper}
A curious observation, somewhat besides the main focus of this paper, 
is that the last Dynkin diagram of the previous
subsection has the same
form as the one for $E_6$, 
had the extra node been white
instead of grey. Similarly, had the extra node in the
superalgebra of Subsection \ref{NullVectorsSub} been white, we would
have had the Lie algebra for $SO(10)$, or in the general case $SO(D+2)$.

Polynomials of a pure spinor, the constrained object encoding the
spectrum of the Borcherds algebra in question, 
indeed form an infinite-dimensional
``singleton'' representation of $E_6$, which can be constructed as
follows.
Consider generators of $E_6\supset\mathfrak{so}(10)\oplus\mathfrak u(1)$. 
The adjoint splits as
${\bf78}\rightarrow\overline{\bf16}_{-1}\oplus({\bf45}\oplus{\bf1})_0
\oplus{\bf16}_1$. Call
the spinorial generators $\lambda^\alpha$ and $\mu_\alpha$. With the
conventions
\begin{align}
[J_{ab},\lambda^\alpha]=\frac14(\gamma_{ab}\lambda)^\alpha
\,,\quad[Q,\lambda^\alpha]=\lambda^\alpha\,,
\end{align}
the only non-manifestly covariant non-vanishing commutator is
\begin{align}
[\mu_\alpha,\lambda^\beta]=(\gamma^{ab})_\alpha{}^\beta J_{ab}
+\frac32\delta_\alpha^\beta Q\,.
\end{align}
The relative coefficient is fixed by demanding the Jacobi identity on
the form 
\begin{align}
[[\mu_\alpha,\lambda^{[\beta}],\lambda^{\gamma]}]=0\,.
\end{align}

Now, we start from an $\mathfrak{so}(10)$-scalar "ground state" $\ket0$
annihilated by $\mu_\alpha$, and 
use $\lambda^\alpha$ as "creation operators", giving a Verma module of
polynomials in $\lambda$. 
Let the ground state
have charge $q$, $Q\ket0=q\ket0$.
We want to adjust the value of $q$ so that $(\lambda\gamma^a\lambda)$
generates an $E_6$-invariant ideal. This happens if 
$\mu_\alpha(\lambda\gamma^a\lambda)\ket0=0$. A short calculation
leads to
\begin{align}
\mu_\alpha(\lambda\gamma^a\lambda)\ket0
&=\frac32(2q+1)(\gamma^a\lambda)_\alpha\ket0
+\frac14(\gamma^{ij}\gamma^a\gamma_{ij}\lambda)_\alpha\ket0\cr
&=\left(\frac32(2q+1)-\frac{27}2\right)(\gamma^a\lambda)_\alpha\ket0\,.
\end{align}
If $q=4$, this vanishes, and the ideal may be factored out without breaking
$E_6$.
This shows how the space of (holomorphic) 
polynomials in a pure spinor 
forms an infinite-dimensional lowest-weight representation of 
$E_6$. 
It may be called a singleton representation, since it
only consists of a ``leading trajectory'' of $Spin(10)$ 
representations with
highest weights $(0000k)$ at each $U(1)$ charge $4+k$.
The lowering operator $\mu_\alpha$ can be identified with the gauge
invariant (in the sense that it respects the ideal) 
derivative with respect to a pure spinor constructed in ref. 
\cite{Cederwall:2010zz}.

In the same way a singleton representation of $SO(D+2)$ 
\cite{Dirac:1963} is obtained by
starting from a ground state of a certain charge.
We use the conventions 
\begin{align}
&[J_{ab},\lambda_c]=2\lambda_{[a}\eta_{b]c}\,,\quad
[Q,\lambda_a]=\lambda_a\,,\cr
&[\mu_a,\lambda_b]= J_{ab}-\eta_{ab}Q\,.
\end{align}
The ideal generated by $\lambda^2$ can be factored out for vacuum
charge $q=\frac{D-2}{2}$, leading to a singleton representation.

At least in these particular cases, the constrained object, {\it i.e.},
the pure spinor or null vector, has a direct relation both to a Borcherds
superalgebra and to a (finite-dimensional) Lie algebra.
Both algebras are obtained by adding a node, grey and white,
respectively, in the same position to the Dynkin diagram of a
semi-simple Lie algebra, but the r\^ole of the pure spinor is quite
different in the two cases.
The phenomenon is certainly more general, but is in its present form limited to
situations where the $\lambda$'s commute, and thus are the only
generators at positive level in the Lie algebra. It holds {\it e.g.} for the
constrained objects $\lambda$ occurring in relation to tensor hierarchies of
$E_{n}\times\mathbb{R}^+$ with $n\leq6$, 
and the Borcherds superalgebras obtained by extending $E_{n}$
by an odd null root. The corresponding Lie algebra, having polynomials
of $\lambda$ as a singleton representation is 
$E_{n+1}\supset E_{n}\times\mathbb{R}^+$, which is a 3-grading.

\vskip1cm
\noindent\underbar{Acknowledgements:} MC would like to thank Sergio
Benvenuti and Amihay Hanany for
pointing out parallels with the work of refs.
\cite{Benvenuti:2010pq,Hanany:2014dia}, where similar partition
functions occur \eg\ in relation to moduli spaces of instantons. 
JP would like to thank Thomas Edlund, Alex Feingold,
Sebastian Guttenberg, Axel Kleinschmidt and Arne Meurman for discussions.
The work of JP is supported by NSF grant PHY-1214344.

\newpage

\bibliographystyle{utphysmod2}

%\bibliography{biblio}

\providecommand{\href}[2]{#2}\begingroup\raggedright\endgroup

\end{document}